\documentclass[preprintnumbers,superscriptaddress,pra,showkeys]{revtex4}
\usepackage{amsmath}
\usepackage{amsfonts}
\usepackage{amssymb}
\usepackage{graphicx}
\usepackage{color}

\begin{document}

\title{Thermodynamic pressure and mechanical pressure for electromagnetic
media}
\author{Q. H. Liu}
\email{quanhuiliu@gmail.com}
\affiliation{School for Theoretical Physics, School of Physics and Electronics, Hunan
University, Changsha 410082, China}
\date{\today }

\begin{abstract}
By the mechanical pressure we mean that the pressure in the fundamental
thermodynamic equation with the naive form of the electromagnetic work used,
while the thermodynamic one we mean that in the equation with proper
thermodynamic form of the electromagnetic work instead. Both pressures
differ from a magnetic mutual field pressure which results from the
electromagnetic stress tensor for the linear and uniform media in static
electromagnetic field. Both pressures are in essence tensors, but a
quasi-scalar theory is sufficient for the simple media.
\end{abstract}

\keywords{Maxwell stress tensor, magnetization, electromagnetic pressure,
mutual field energy, electromagnetic energy.}
\maketitle

\section{Introduction}

There are some theoretical puzzles in thermodynamics in the presence of
electromagnetic fields, both in pedagogical/elementary \cite%
{Mandl,Callen,teaching,books,MFE} in research/advanced \cite%
{exp,research,mmfp} aspects. It is well-known that for a paramagnetic
medium, at sufficiently high magnetic fields all magnetic domains will
expand to their maximum size and/or rotate in the direction of external
field. \cite{web} It indicates that once the solid medium is carefully
prepared, both magnetostriction and the piezomagnetism is significant along
the direction of the magnetic field, the thermodynamic effects coupling the
geometry of the medium and applied field. To account for such effects, we
must introduce the tensor analysis to the fundamental thermodynamic
equation, even for the simplest sample, the solid media that is linear,
uniform and isotropic in nature. However, the usual quasi-scalar theory
suffices for the simplest medium, with some clarifications. 

It is beneficial to distinguish two forms of the electromagnetic work
element: One is $Ydy$ where $\left( Y,y\right) $ are a conjugate pair in
energy representation, and $Y$ and $y$ are intensive and extensive quantity,
respectively, \cite{books,Callen} and another is $ydY$. \cite{books,MFE} The
difference in between is usually overlooked or simply treated. However, some
insists that only the former is correct \cite{books} while some prefers the
latter. \cite{MFE} We explicitly treat the elastic and solid paramagnetic
materials and the equilibrium thermodynamics and reversible processes. Our
results are applicable for dielectrics as well, with a simple replacement of
the symbols, which will not be explicitly treated. 

The energy density element for the linear and uniform magnetic media in
static magnetic field is%
\begin{equation}
d\omega =\mathbf{H\cdot }d\mathbf{B}=HdB  \label{01}
\end{equation}%
where $\mathbf{H}$ and $\mathbf{B}$ symbolize the magnetic strength and
magnetic induction, respectively, and $\mathbf{B=}\mu _{0}\left( \mathbf{H+M}%
\right) $ with $\mu _{0}$ being the vacuum permeability and $\mathbf{M}$
being the magnetization strength. Assume that the volume of the magnetic
material is $V$, and the energy element d $W_{B}^{me}$ is from \emph{%
electromagnetism} \cite{books,MFE}%
\begin{equation}
dW_{B}^{me}=\int_{V}d\omega dV=\mu _{0}\int_{V}HdHdV+\mu
_{0}\int_{V}HdMdV=\mu _{0}Vd\frac{H^{2}}{2}+\mu _{0}VHdM.  \label{02}
\end{equation}
To note that the first part $\mu _{0}VdH^{2}/2$ in the work element can be
simply removed for it is uniformly distributed in whole space which has no
thermal consequences. Thus the mechanical energy work element is%
\begin{equation}
dW_{B}^{me}=\mu _{0}VHdM=\mu _{0}Hdm-\mu _{0}HMdV=d%
W_{B}^{th}-\mu _{0}HMdV,  \label{03}
\end{equation}%
in which the first part is defined as thermodynamic energy density element%
\begin{equation}
dW_{B}^{th}=\mu _{0}\mathbf{H\cdot }d\mathbf{m=}\mu _{0}Hdm
\label{04}
\end{equation}%
where the total magnetic moment $\mathbf{m}$ is%
\begin{equation}
\mathbf{m}=\mathbf{M}V.  \label{05}
\end{equation}%
Why $\mu _{0}Hdm$ (\ref{04}) can be termed as the thermodynamic energy
density element d $W_{B}^{th}$ is due to a fact the axiomatic formalism of
the thermodynamics assumes that the work element takes the form $Ydy$, \cite%
{Callen} and only this form can give the correct form of the experimental
results for relationship between magnetostriction and the piezomagnetism. 
\cite{exp} 

For a fixed number of molecules for the simple magnetic media, the
fundamental thermodynamic equation is 
\begin{equation}
dU=TdS-p_{th}dV+%
\mu
_{0}Hdm  \label{fte1}
\end{equation}%
where $S$ is entropy. Here, the pressure $p_{th}$ is an \emph{external} and 
\emph{total} quantity causing the volume change $-dV$. Both pressure $p_{th}$
and work $%
\mu
_{0}Hdm$ (\ref{03}) are thermodynamic. To note that the equation (\ref{fte1}%
) can transformed into 
\begin{equation}
dU=TdS-p_{me}dV+\mu _{0}VHdM=TdS-\left( p_{me}+%
\mu
_{0}HM\right) dV+%
\mu
_{0}Hdm  \label{fte2}
\end{equation}%
The relation between two pressures $p_{th}$ and $p_{me}$ is 
\begin{equation}
p_{th}=p_{me}+%
\mu
_{0}HM  \label{twop}
\end{equation}%
where $%
\mu
_{0}HM$ are magnetic mutual field pressure. \cite{mmfp} 

The main aim of the present paper is three-fold: 1) To demonstrate that this
magnetic mutual field pressure $%
\mu
_{0}HM$ is a reasonable consequence of the proper form of the
electromagnetic stress tensor, a mechanical interaction between the magnetic
field and material, 2) to show that both forms of pressure are legitimate
but applicable with different form of energy density, 3) to illustrate that
both pressures are in essence tensors, but the quasi-scalar form suffices
provided that the sample is specially prepared, with a critical comment on 
\emph{seemingly reasonable} results (\ref{03})-(\ref{twop}). 

The paper is organized in the following. Section II shows how the magnetic
mutual field pressure naturally appears in the Maxwell stress tensor as a
component of the tensor. Section III is a brief conclusion and discussion.

\section{Magnetic mutual field pressure and the Maxwell stress tensor}

When the media are static rather than moving, the Einstein and Laub form of
the Maxwell stress tensor \cite{EL} is proper, which has been deeply
understood recently as a by-product of the intensive exploration of its
application to clarify the famous Abraham-Minkowski debate, \cite{el1} and
other problems. \cite{el2} The Einstein and Laub tensor is \cite{Einstein}%
\begin{equation}
T_{ij}=E_{i}D_{j}+H_{i}B_{j}-\frac{1}{2}(\epsilon _{0}E^{2}+%
\mu
_{0}H^{2})\delta _{ij}=E_{i}D_{j}+H_{i}B_{j}-u\delta _{ij}  \label{e}
\end{equation}%
which for magnetic field becomes%
\begin{equation}
T_{ij}=H_{i}B_{j}-\frac{1}{2}%
\mu
_{0}H^{2}\delta _{ij}=H_{i}B_{j}-u\delta _{ij}.
\end{equation}%
Here $\epsilon _{0}$ and $%
\mu
_{0}$ are the electric and magnetic constants, respectively, and $\delta
_{ij}$ is the Kronecker delta, and $\mathbf{E}$ the electric field, $\mathbf{%
H}$ the magnetic field, and electric displacement $\mathbf{D=}\epsilon _{0}%
\mathbf{E}+\mathbf{P}$ with $\mathbf{P}$ the polarization and magnetic
induction $\mathbf{B=}$ $%
\mu
_{0}\left( \mathbf{H}+\mathbf{M}\right) $ with $\mathbf{M}$ the
magnetization; and $u=\left( \epsilon _{0}E^{2}+%
\mu
_{0}H^{2}\right) /2$ is the energy density in vacuum which is $%
\mu
_{0}H^{2}/2$ with magnetic field only.

For simplicity, the medium is cylindrical in shape and is placed inside a
solenoid coaxial to the sample. \cite{Callen} A current in the solenoid is
gradually switched on, and a uniform axial magnetic field is built inside
the solenoid to magnetize the sample. We assume that the magnetic field is
along the $z$-axis, see Fig. 1. To determine the pressure tensor, we use two
boxes 1 and 2 in Fig. 1, and we have two tensors $T^{inner}$ and $T^{outer}$
for the inner and the outer part of two boxes across the surface,
respectively%
\begin{equation}
T^{inner}=\left( 
\begin{array}{ccc}
-u & 0 & 0 \\ 
0 & -u & 0 \\ 
0 & 0 & u+%
\mu
_{0}HM%
\end{array}%
\right) ,T^{outer}=\left( 
\begin{array}{ccc}
-u & 0 & 0 \\ 
0 & -u & 0 \\ 
0 & 0 & u%
\end{array}%
\right) .
\end{equation}%
To compute the components of force $\Delta \mathbf{F}$ in $x$ and $y$
direction, we need to use the box 1. We have $\Delta F_{x}=\Delta F_{y}=0$.
To know the component of force $\Delta \mathbf{F}$ in $z$ direction, we need
to use box 2, and the result is, 
\begin{equation}
\Delta F_{z}=\sum_{j=1}^{3}\oint T_{zj}ds_{j}=0+0+\oint T_{zz}ds_{z}=\left(
T^{outer}\right) _{zz}\Delta s_{z}+\left( T^{inner}\right) _{zz}(-\Delta
s_{z})=-%
\mu
_{0}HM\Delta s_{z}
\end{equation}%
where $\Delta s_{z}$ is the area of one small surface of two sides of box 2
parallel to the $z$-axis. We have the components of the magnetic mutual
field pressure tensor%
\begin{equation}
p_{ij}=\left\{ 
\begin{array}{c}
-%
\mu
_{0}HM \\ 
0%
\end{array}%
,%
\begin{array}{c}
i=j=3 \\ 
otherwise%
\end{array}%
\right. .  \label{p}
\end{equation}%
Here the pressure tensor has nonvanishing effect along the direction of
external field only. Assume that there is still mechanical pressure tensor $%
\left( p_{me}\right) $ inside the material without the field%
\begin{equation}
\left( p_{me}\right) =\left( 
\begin{array}{ccc}
p_{me,x} & 0 & 0 \\ 
0 & p_{me,y} & 0 \\ 
0 & 0 & p_{me,z}%
\end{array}%
\right) ,
\end{equation}
in which three components $p_{me,i}$ ($i=x,y,z$) are in general not equal to
each other, we have explicitly thermodynamic pressure tensor $\left(
p_{th}\right) $  
\begin{equation}
\left( p_{th}\right) =\left( 
\begin{array}{ccc}
p_{me,x} & 0 & 0 \\ 
0 & p_{me,y} & 0 \\ 
0 & 0 & p_{me,z}+%
\mu
_{0}HM%
\end{array}%
\right) .
\end{equation}

So far we see clearly, usual mechanic pressure $p_{me}$ is isotropic for
fluid media   
\begin{equation}
\left( p_{me}\right) =\left( 
\begin{array}{ccc}
p_{me} & 0 & 0 \\ 
0 & p_{me} & 0 \\ 
0 & 0 & p_{me}%
\end{array}%
\right) .
\end{equation}%
For the isotropic fluid medium in a circular cylinder of cross-section $S$,
we can control volume compression/expansion with a piston such that the work 
$p_{me}dV$ is meaningful in the following sense $p_{me}Sdl$ where $dl$ is
the axial displacement of the piston. When an external field $\mathbf{H=}H%
\mathbf{e}_{z}$ is applied with $\mathbf{e}_{z}$ axially pointing, the
mechanic pressure $p_{me}$ inside the sample can hardly be held isotropic
for there is the magnetic mutual field pressure $%
\mu
_{0}HM$ is only along the direction of the magnetic field. In other words,
there is an axial strain of the media due to the presence of the magnetic
field. Thus a precise understanding of the results (\ref{03})-(\ref{twop})
is given in the following. When the magnetic media is also circular cylinder
in shape and the external field is along the axial direction as shown in
Fig. 1, the volume change is $dV=Sdz$ in a response of both the magnetic
mutual field pressure $%
\mu
_{0}HM$ and external pressure $p_{th}$ 
\begin{equation}
\mathbf{F\cdot }d\mathbf{z=}p_{th}Sdz=\left( p_{me}+%
\mu
_{0}HM\right) Sdz.
\end{equation}
Once the $H$ reduces to zero, $p_{me}$ must increase to balance the pressure 
$p_{th}$.

\section{Discussion and conclusion}

We start from introduction of mechanical pressure and energy density in the
thermodynamic fundamental equations, and reach the thermodynamic ones which
is indicated by axiomatic formulation of the thermodynamics, with an
anisotropic magnetic mutual field pressure be introduced. Both pressures are
in essence tensors, but the quasi-scalar form suffices provided that the
sample is specially prepared. 

\begin{acknowledgments}
This work is financially supported by the Hunan Province Education Reform
Project under Grants No. HNJG-2022-0506 and No. HNJG-2023-0147. The author
is indebted to Professor Xiaofeng Jin at Fudan University, and the members
of Online Club Nanothermodynamica (Founded in June 2020), and members of
National Association of Thermodynamics and Statistical Physics Teachers in
China, for fruitful discussions.
\end{acknowledgments}

\begin{figure}[h]
\includegraphics[height=9cm]{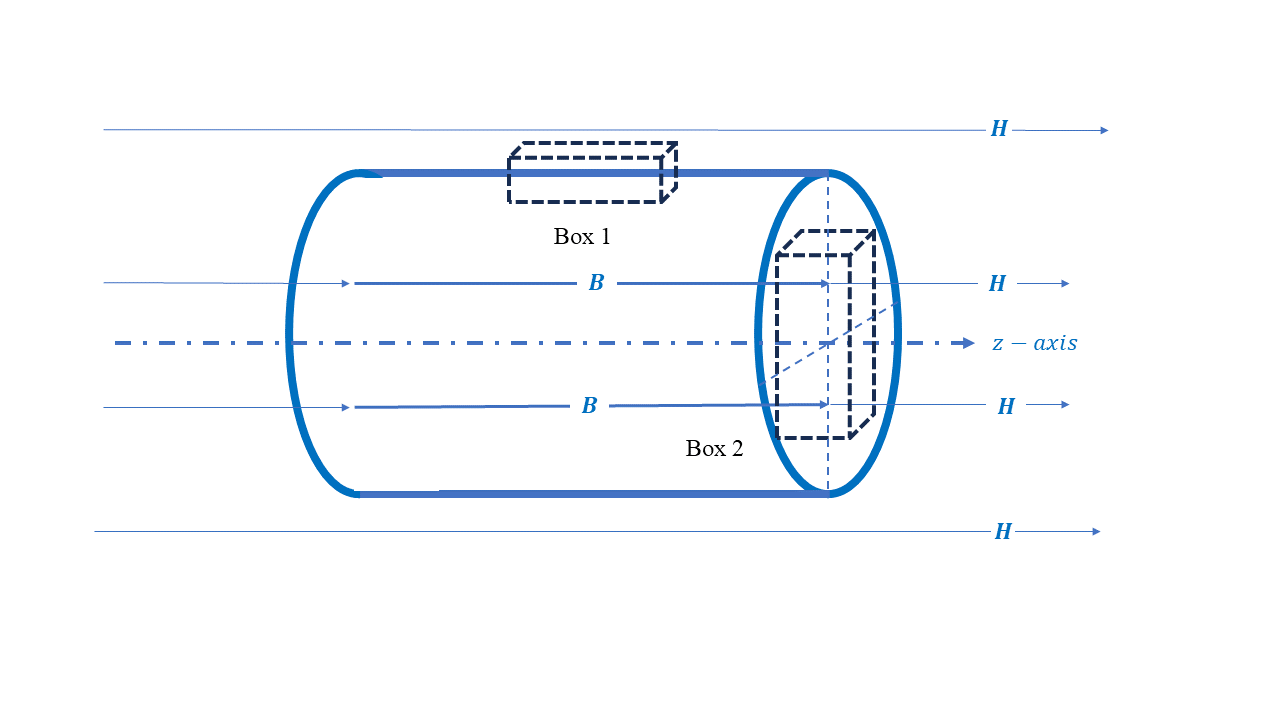}
\caption{The magnetic media inside the solenoid is sketched, which is
uniformly magnetized. Two small boxes cross the surfaces of the media are
used to calculate the pressure tensor near the surface.}
\label{Fig1}
\end{figure}

\end{document}